# Diffusion Kurtosis Imaging maps neural damage in the EAE model of multiple sclerosis

Andrey Chuhutin[1], Brian Hansen[1], Agnieszka Wlodarczyk[2], Trevor Owens[2], Noam Shemesh[3], Sune Nørhøj Jespersen[1,4]

**Abstract**

Diffusion kurtosis imaging (DKI), is an imaging modality that yields novel disease biomarkers and in combination with nervous tissue modeling, provides access to microstructural parameters. Recently, DKI and subsequent estimation of microstructural model parameters has been used for assessment of tissue changes in neurodegenerative diseases and their animal models. In this study, mouse spinal cords from the experimental autoimmune encephalomyelitis (EAE) model of multiple sclerosis (MS) were investigated for the first time using DKI in combination with biophysical modeling to study the relationship between microstructural metrics and degree of animal dysfunction. Thirteen spinal cords were extracted from animals of variable disability and scanned in a high-field MRI scanner along with five control specimen. Diffusion weighted data were acquired together with high resolution $T_2^*$ images. Diffusion data were fit to estimate diffusion and kurtosis tensors and white matter modeling parameters, which were all used for subsequent statistical analysis using a linear mixed effects model. $T_2^*$ images were used to delineate focal demyelination/inflammation. Our results unveil a strong relationship between disability and measured microstructural parameters in normal appearing white matter and gray matter. Relationships between disability and mean of the kurtosis tensor, radial kurtosis, radial diffusivity were similar to what has been found in other hypomyelinating MS models, and in patients. However, the changes we found in biophysical modeling parameters and in particular in extra-axonal axial diffusivity were clearly different from previous studies employing other animal models of MS. In conclusion, our data suggest that DKI and microstructural modeling can provide a unique contrast capable of detecting EAE-specific changes correlating with clinical disability. These findings could close the gap between MRI findings and clini-

1    CFIN, Aarhus University, Aarhus, Denmark
2    Department of Neurobiology Research, Institute for Molecular Medicine, University of South Denmark, Odense, Denmark
3    Champalimaud Research, Champalimaud Centre for the Unknown, Lisbon, Portugal
4    Department of Physics, Aarhus University, Aarhus, Denmark

cal presentation in patients and deepen our understanding of EAE and the MS mechanisms.

# I. Abbreviations

| | |
|---|---|
| $D_a$ | intra-axonal diffusivity |
| $D_\perp$ | radial diffusivity |
| $D_\parallel$ | axial diffusivity |
| $K_\perp$ | radial kurtosis |
| $D_{e,\parallel}$ | extra-axonal axial diffusivity |
| $\kappa$ | concentration parameter of the Watson distribution |
| $K_\parallel$ | axial kurtosis |
| $D_{e,\perp}$ | extra-axonal radial diffusivity |
| $f$ | volume fraction of axonal compartment , |
| ANOVA | analysis of variance |
| BW | receiver bandwidth |
| CNS | central nervous system |
| DKI | diffusion kurtosis imaging |
| DTI | diffusion tensor imaging |
| DWI | diffusion weighted imaging |
| EAE | experimental autoimmune encephalomyelitis |
| ESP | echo spacing |
| FA | fractional anisotropy |
| FDR | false discovery rate |
| fODF | fiber orientation distribution function |
| FOV | field of view |
| GM | gray matter |
| LME | linear mixed effects modeling |
| LTO | lower thoracic segment |
| LU | lumbar segment |
| MD | mean diffusivity |
| MKT | mean of the kurtosis tensor |
| MRI | magnetic resonance imaging |
| MS | multiple sclerosis |
| MTO | mid thoracic segment |
| NA | number of averages |
| NAWM | normal appearing white matter |
| NODDI | neurite orientation dispersion and density imaging |
| PBS | phosphate buffered saline |
| PFA | paraformaldehyde |
| RRMS | relapsing-remitting multiple sclerosis |
| SC | spinal cord |
| SEM | scanning electron microscope |
| SMT | spherical means technique |
| TE | echo time |
| TR | repetition time |
| WM | white matter |
| WMM | white matter model |

## II. Introduction

Multiple sclerosis (MS) is a demyelinating, inflammatory, neurodegenerative disease of the human central nervous system (CNS) affecting millions of people worldwide. The pathophysiology of MS is often complex, and involves, among other factors, myelin loss, axonal damage, appearance of transient or permanent lesions, and brain atrophy. Effective treatment of MS is still lacking (Compston and Coles, 2002), although a range of disease-modifying therapies have been introduced (Berger, 2011; Noyes and Weinstock-Guttman, 2013). These therapies are based on immunomodulatory, anti-inflammatory, and immunosuppressive drugs. The success of such treatments depends on early (preferably noninvasive) diagnosis and careful monitoring of the patient.

A range of MS animal models characterized by different mechanisms of induction and pathology (Lassmann and Bradl, 2016) have been developed to overcome the limitations of clinical tissue assessment. Experimental autoimmune encephalomyelitis (EAE) is one of the most compelling and commonly used groups of animal MS models (Baker and Amor, 2014; Kipp et al., 2016; Lassmann and Bradl, 2016). Unlike other animal models, in addition to inflammatory lesions and demyelination, EAE includes salient axonal damage (Bergers et al., 2002; Kipp et al., 2016) which is one of the hallmarks of MS. Therefore, using EAE to assess MS biomarkers can provide advantageous insights into the MS pathology.

Due to its noninvasiveness and ability to contrast soft tissues, Magnetic Resonance Imaging (MRI) is extensively used for diagnosis and monitoring of MS (Bakshi et al., 2008; Polman et al., 2011). Standard $T_1$- or $T_2$-weighted MRI images are capable of revealing brain atrophy and lesions, which are heterogeneous areas harboring demyelination, inflammation, gliosis and axonal in-jury (Filippi et al., 2012; Inglese and Bester, 2010). However, methods based on $T_1$- or $T_2$-weighted maps do not provide the full picture. Diffuse microstructural changes outside the $T_1$ or $T_2$-map intensity lesions in gray matter (GM) and so-called normal appearing white matter (NAWM) (Allen et al., 2001) have been observed with histology. Studies (De Stefano et al., 2006; Kipp et al., 2016; Miller et al., 2003) showed that the diffuse damage in in NAWM and GM contributes to disability accumulation and chronic disease progression.

Diffusion weighted imaging (DWI) can provide quantitative microstructural information by sensitizing MRI signals to the displacement of water molecules. The underlying signal attenuation is often approximated by a Gaussian distribution, which forms the basis of diffusion tensor imaging, DTI (Basser and Pierpaoli, 1996). While DTI metrics are widely used, the mentioned Gaussian approximation is valid only in a limited regime of low diffusion weighting. At higher gradient strengths (or $b$-values), tissue microstructure and compartmentalization increasingly impact the signal and cause deviations from Gaussianity. These deviations are utilized by a framework known as diffusion kurtosis imaging (DKI) (Jensen et al., 2005; Jensen and Helpern, 2010). Combined with tissue modeling, DKI provides access to microstructural parameters and yields novel disease biomarkers.

DKI based biomarkers have been shown to improve the diagnostic assessment in a range of neurological disorders (Delgado y Palacios et al., 2014; Grossman et al., 2011; Khan, 2016; Surova et al., 2016; Tietze et al., 2015; Wang et al., 2011). In MS, they improved characterization of GM and NAWM damage (Raz et al., 2013; Yoshida et al., 2013) and correlated with cognitive impairment (Bester et al., 2015). In MS animal models DKI biomarkers were associated with chronic injury (Falangola et al., 2014; Guglielmetti et al., 2016; Jelescu et al., 2016) and neurite myelin content (Kelm et al., 2016). However, even though (Wu and Cheung, 2010) employed EAE to show that DKI is able to enhance lesion detection and other DWI methods revealed pathological changes in EAE (Biton et al., 2005; Budde et al., 2009), DKI and WM models have never been used to investigate EAE-induced disability.

In this study we hypothesized that novel metrics obtained using DKI could provide a diagnostic tool for MS. Therefore, we explored their possible relationship to EAE disability. A detailed description of the chosen DKI-derived metrics is provided in the next chapter following the definition of the metrics and a description of their relationship to the pathology.

The ability of DKI to provide quantitative biomarkers of dysfunction in EAE model of MS was investigated by testing the correlation between the biomarkers and behavioral markers of disease severity. Inside lesions no biomarkers showed correlation to disability. In NAWM, the DKI parameters showed better correlation to disability than DTI, suggesting that changes in kurtosis parameters may precede lesion formation. Standard DKI and DTI parameters produced results similar to those shown previously in other MS models. The estimated parameters of the white matter model, however, yielded new microstructural information that could provide a key for improved understanding of EAE mechanisms.

## III. Methods

### a. Theory. Diffusion Kurtosis Imaging

DKI (Jensen et al., 2005) improves the approximation of the diffusion weighted signal in vivo (Filli et al., 2014; Raz et al., 2013; Rosenkrantz et al., 2015) and ex vivo (Veraart et al., 2011) by including the next term in the cumulant expansion (Kiselev, 2010; van Kampen, 2007) of the DWI signal $S$

$$\log S(b, \hat{\mathbf{n}}) = -b \sum_{i,j=1}^{3} D_{i,j} \hat{\mathbf{n}}_i \hat{\mathbf{n}}_j + \frac{b^2 \bar{D}^2}{6} \sum_{i,j,k,l=1}^{3} W_{i,j,k,l} \hat{\mathbf{n}}_i \hat{\mathbf{n}}_j \hat{\mathbf{n}}_k \hat{\mathbf{n}}_l \qquad \text{(Equation 1)}$$

where $D_{i,j}$ is the $i,j$ element of the rank 2 symmetric diffusion tensor $\mathbf{D}$ and $W_{i,j,k,l}$ is the $i,j,k,l$ element of the symmetric rank 4 kurtosis tensor $\mathbf{W}$, $b$ is the diffusion weighting ($b$-value), $\bar{D} = \text{Tr}(\mathbf{D})/3$ and $\hat{\mathbf{n}}_i$ denotes the $i$-th component of measurement direction $\hat{\mathbf{n}}$. In analogy to diffusion tensor based fractional anisotropy (FA), mean (MD), axial ($D_\parallel$) and radial ($D_\perp$) diffusivity, the kurtosis tensor ($\mathbf{W}$) provides additional biomarkers: kurtosis fractional anisotropy (KFA) (Hansen and Jespersen, 2016), mean of the kurtosis tensor (Hansen et al., 2014, 2013)(MKT = $\frac{\text{Tr}(\mathbf{W})}{5}$), axial ($K_\parallel$) and radial kurtosis ($K_\perp$) (Jensen and Helpern, 2010).

The choice of the DTI parameters assessed in this study was based on the previous works. In particular, FA, MD, $D_\parallel$ and $D_\perp$ have been shown to be affected by MS pathology (Ceccarelli et al., 2007; de Kouchkovsky et al., 2016; Falangola et al., 2014; Guglielmetti et al., 2016; Inglese and Bester, 2010; Jelescu et al., 2016; Kelm et al., 2016; Mesaros et al., 2009). In DKI, a choice of MKT was motivated by a decrease in mean kurtosis that was shown in human GM and NAWM (Raz et al., 2013; Yoshida et al., 2013) and linked with cognitive impairment in MS (Bester et al., 2015). A decrease in axial kurtosis $K_\parallel$ and radial kurtosis $K_\perp$ was detected in an animal model of chronic MS (Falangola et al., 2014; Guglielmetti et al., 2016), $K_\perp$ was found to be related to the myelin content (Kelm et al., 2016).

In this work we used a variant of the 'standard' WM model (WMM) that has been extensively explored recently (Fieremans et al., 2011; Jelescu et al., 2015a; Jespersen et al., 2007; Novikov et al., 2018; Zhang et al., 2012). The model is designed to approximate diffusion inside and outside WM fascicles in SC. It consists of two non-exchanging Gaussian compartments representing extra-axonal and intra-axonal space. Here, the diffusion in the extra-axonal space is approximated by a tensor which is characterized by extra-axonal radial and axial diffusivities ($D_{e,\perp}$ and $D_{e,\parallel}$). Axons, having radii much smaller than the diffusion distance, are assumed to appear as one-dimensional sticks and thus only the intra-axonal axial diffusivity $D_a$ is non-vanishing. Taking $f$ to be the volume fraction of the axonal compartment, and $\mathcal{P}(\hat{\mathbf{u}})$ to be the fiber-orientation distribution function (fODF), the diffusion signal $S$ measured in the direction $\hat{\mathbf{n}}$ can be written as:

$$S(b, \hat{\mathbf{n}}) = \int d\hat{\mathbf{u}} \mathcal{P}(\hat{\mathbf{u}}) \left( f \exp\left(-bD_a (\hat{\mathbf{u}} \cdot \hat{\mathbf{n}})^2\right) \right.$$
$$\left. + (1-f) \exp\left(-bD_{e,\perp} - b\left(D_{e,\parallel} - D_{e,\perp}\right)(\hat{\mathbf{u}} \cdot \hat{\mathbf{n}})^2\right) \right) \quad \text{(Equation 2)}$$

This formulation assumes fODF having a special axially-symmetric form (e.g., the Watson distribution $\mathcal{P}(\hat{\mathbf{u}}) \propto \exp\left(\kappa (\hat{\mathbf{u}} \cdot \hat{\mathbf{c}})^2\right)$, where $\kappa$ is the concentration parameter and $\hat{\mathbf{c}}$ is the symmetry axis. Therefore it does not share the typical assumption of parallel fibers (Fieremans et al., 2011). The relationship between the parameters of this WMM and elements of $\mathbf{W}$ and $\mathbf{D}$ can be established (Novikov et al., 2018), (Jespersen et al., 2017) and the corresponding model parameters are expected to be valid for more general tissue types than in previous studies (de Kouchkovsky et al., 2016; Falangola et al., 2014; Jelescu et al., 2016).

For this study the WMM parameters chosen to be assessed were those sensitive to neural damage (Falangola et al., 2014; Kelm et al., 2016); in particular, $f$, a biomarker for axonal loss (Fieremans et al., 2012) linked to myelin content and axon density (Kelm et al., 2016). $D_a$ which is associated with intra-axonal injury (Hui et al., 2012b), $D_{e,\perp}$ which is related to the g-ratio (Jelescu et al., 2016) and $D_{e,\parallel}$ which is a marker of demyelination (Fieremans et al., 2012) through tortuosity. In addition, $\kappa$ that reflects the fiber dispersion (Grussu et al., 2017) was studied.

For this study DKI data was used as a starting point for WMM parameters estimation. This approach is commonly used in different models of tissue microstructure (Fieremans et al., 2011; Hansen et al., 2017; Hui et al., 2015; Jespersen et al., 2012; Novikov et al., 2018; Novikov and Kiselev, 2010; Szczepankiewicz et al., 2016). Fitting DKI enables usage of linear least squares algorithms that yield stable estimates (Chuhutin et al., 2017) decreasing the chances to end up in a local minimum. Moreover, WMM fit at order $b^2$ yields two solutions that fit data equally well (Novikov et al., 2018). Using DKI fit and consequently estimating the WMM parameters allows to choose a particular solution branch (Jelescu et al., 2015b) explicitly.

### b. Animal treatment

Female C57BL/6j bom (B6) mice aged 6 to 8 weeks obtained from Taconic Europe A/S, (Lille Skensved, Denmark) were maintained in the Biomedical Laboratory, University of Southern Denmark (Odense).

Mice were immunized by injecting subcutaneously 100μl of an emulsion containing 100μg myelin oligodendrocyte glycoprotein (MOG)$_{p35-55}$ (TAG Copenhagen A/S, Frederiksberg, Denmark) in incomplete Freund's adjuvant (DIFCO, Aberstslund, Denmark) supplemented with 400 µg H37Ra *Mycobacterium tuberculosis* (DIFCO). *Bordetella pertussis* toxin (300 ng; Sigma-Aldrich, Brøndby, Denmark) in 200 µl of PBS was injected intraperitoneally at day 0 and day 2. Animals were monitored daily from day 5 and scored on a 6-point scale as follows: 0, no symptoms; 1, partial loss of tail tonus; 2, complete loss of tail tonus; 3, difficulty walking; 4, paresis in both hind legs; 5, paralysis in both hind legs; and 6, front limb weakness. About 75% of the mice showed symptoms of EAE. All the scoring was performed by the same person (AW) with previous experience of EAE animal assessment (Wlodarczyk et al., 2014). Severe EAE usually developed 14 to 18 days after immunization. Based on the provided EAE-scale, the animals were divided into roughly equisized groups of samples: low-grade (EAE score 1.5-2, 5 samples), intermediate (2.5-4, 3 samples), high (4.5-5, 5 samples). If not stated otherwise, the control group is henceforth referred to as zero-grade for convenience (5 samples).

Animal experiments were approved by Danish Animal Experiments Inspectorate (approval number 2014-15-0201-00369).

### c. Sample preparation

Mice were euthanized by pentobarbytol overdose and transcardially perfused with PBS followed by 4% buffered paraformaldehyde (PFA) (pH 7.4). The spinal column was extracted and stored in 4% PFA for 7 days. On day 8, now fully fixed cords were manually dissected out of spinal column, and stored in 4% PFA with the meninges removed until MRI. 24 hours prior to the experiment the samples were washed in phosphate buffered saline (PBS) to remove PFA and to minimize associated $T_2^*$-related signal attenuation (Shepherd et al., 2009, 2005). The SC was cut into 3 parts and segments from T8 up to L6 were selected for imaging. We differentiate between three segments of mouse SC as follows: mid-thoracic (MTO):T8-T11, lower thoracic (LTO):T12-LU1, lumbar (LU):L2-L6.

### d. MR imaging

Imaging was performed on a 16.4 T vertical bore Bruker Biospin (Ettlingen, Germany) Aeon Ascend magnet equipped with a Micro5 probe and a gradient unit capable of delivering up to 3000 mT/m in all directions. The samples were placed in a 5 mm NMR tube filled with Fluorinert© 3M and held parallel to the direction of the main magnetic field using a polypropylene straw. The temperature was monitored and maintained at 23.6ºC using air flow.

Diffusion kurtosis data were acquired using a 2D diffusion weighted fast spin echo sequence with echo train length (ETL=8), first echo time (effective TE = 15 ms), echo spacing (ESP=4.23 ms), total repetition time (TR=2000ms) (Beaulieu et al., 1993; Kelm et al., 2016; West et al., 2018). Receiver band-width (BW) for signal acquisition = 83 kHz. For each SC, between 16 and 22 0.5mm-thick slices were scanned. For each slice, a matrix of size 120x120 voxels with field of view (FOV = 4.2 mm x 4.2 mm) (resolution of 0.035 mm x 0.035 mm) was acquired. Diffusion weighting was performed with short gradient pulses of duration ($\delta$ =1.5 ms) and separation (diffusion time) $\Delta = 10$ms. Diffusion weighting with $b$-values of 0.2,0.3,0.5,0.6,0.9,1,1.2,1.5,1.8,2.1,2.5 ms/µm$^2$ was applied along 30 directions, with 1 average (NA) for $b$<1.2 ms/µm$^2$ and 2 averages for $b$>1.2 ms/µm$^2$. Sixty $b$=0 ms/µm$^2$ images were collected for normalization. The total scanning time per spinal cord was about 10 hours. Examples of images acquired with $b = 0.2\,\text{ms}\,\mu\text{m}^{-2}$ are provided in Fig. 1 (A,B). SNR (amplitude ratio) of the acquired raw data was estimated to be ~30-40 (in WM, $b$=0), ~50 (in GM, $b$=0), ~30 (in WM, $b = b_{\max}$), ~20 (in GM, $b = b_{\max}$).

High resolution $T_2^*$-weighted images for lesion delineation were acquired using fast low angle shot (FLASH) pulse sequence with twice the in-plane resolution (0.018 mm x 0.018 mm) and the same slice thickness (0.5mm), NA=2 and TE = 5 ms.

**e. Image segmentation**

Image segmentation of white and gray matter was performed manually based on the mouse spinal cord atlas (Watson, 2009).

Lesions were manually outlined on $T_2^*$-weighted slices as described in (Steinbrecher et al., 2005) and thereafter lesion maps were downsampled to the resolution of DWI maps. On each slice, potential abnormalities were inspected and compared to the atlas. Voxels with abnormal hyperintensity that could not be explained by the anatomical features of SC, were manually marked using an in-house developed software tool. Delineation followed a conservative definition of the lesion. As such, whenever when there was a suspicion that the increase in WM intensity could be explained by anatomical features, the voxels were not delineated as lesion. The slices and spinal cords were presented in randomized order and the examiner (AC) was blinded to the grade. An example of this segmentation is shown in Fig. 2. NAWM was defined after the segmentation as a non-lesion WM. Lesion load was defined as fraction of volume taken by abnormal hyperintensity in $T_2^*$-maps.

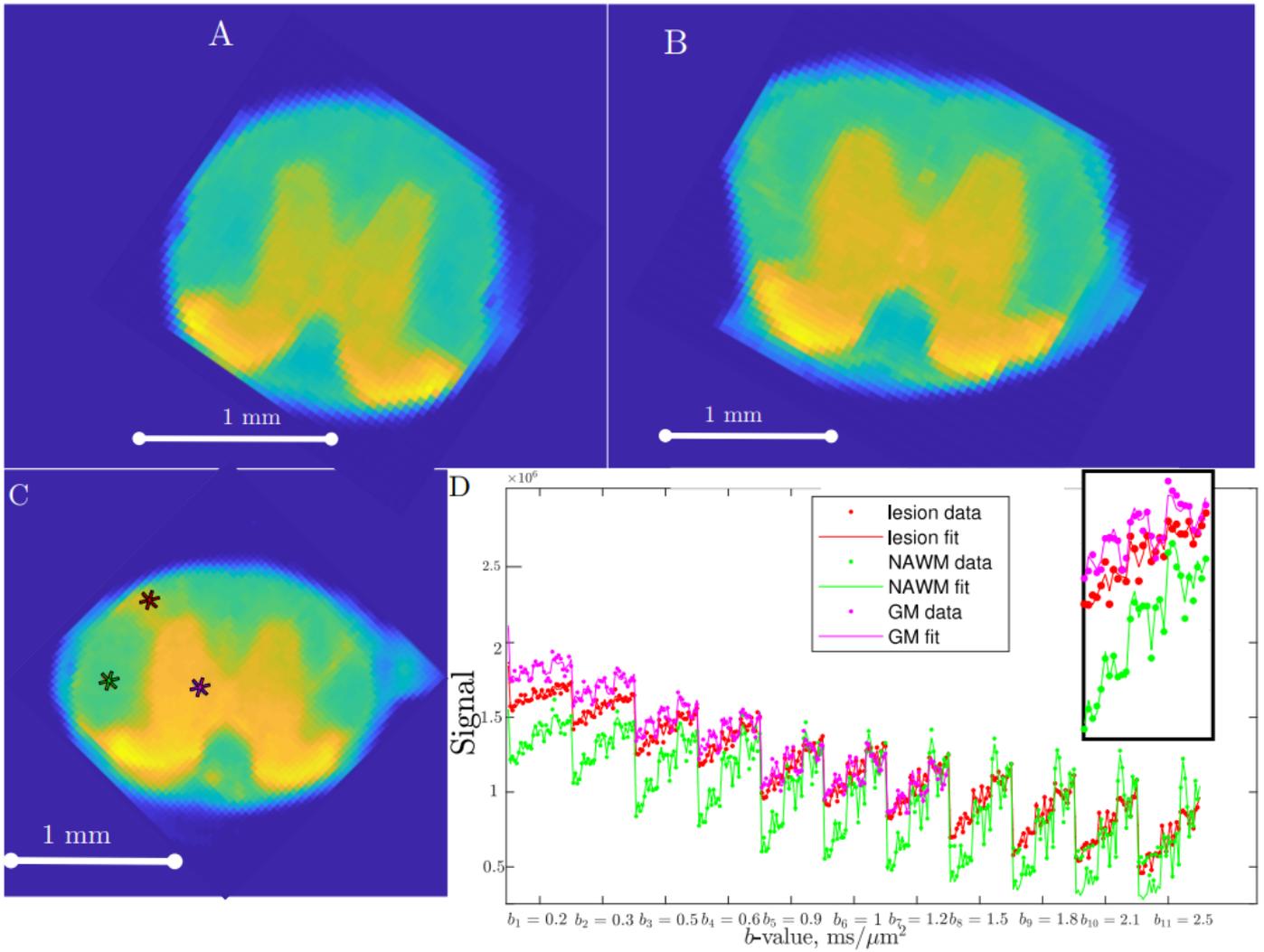

Figure 1: Example of acquired raw data and a corresponding data fit. Subfigures (A) and (B) show an example of a raw signal image acquired for low diffusion weighting ($b=0.2\mu mms^{-2}$) in mid thoracic and low thoracic segments of a control sample. Subfigure (C) shows a high grade sample, with a visible lesion acquired with the same low diffusion weighting ($b=0.2\mu mms^{-2}$). Data and data fit that correspond to three different voxels in the slice denoted in (C) are shown in subplot (D). Lesion voxel location is marked in red, NAWM voxel in green and GM voxel in magenta in subplot (C). Multiple data points plotted under each b-value on the x-axis correspond to different directions. The inset shows the part of the graph corresponding to $b=0.5$ $\mu mms^{-2}$ enlarged.

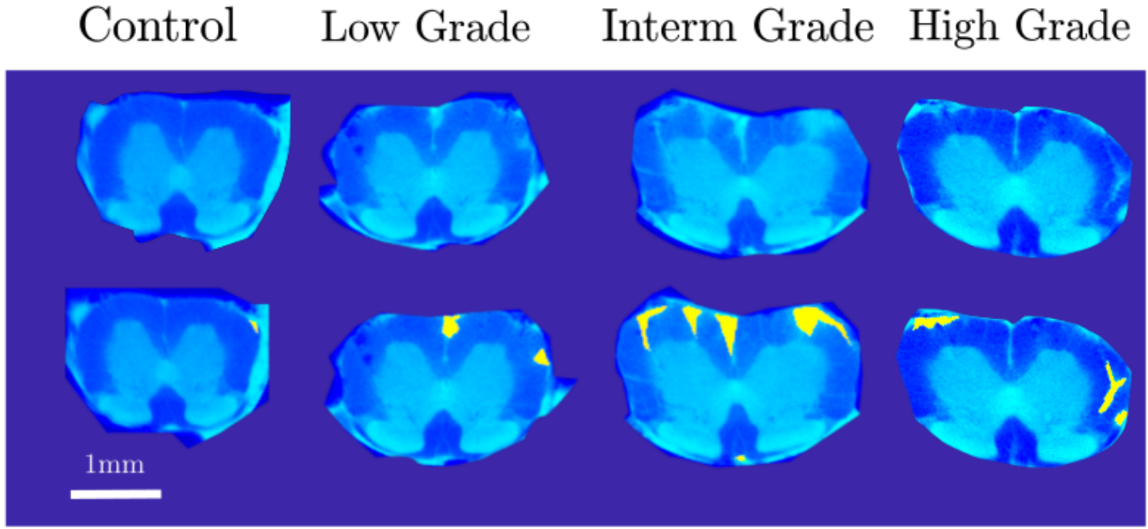

*Figure 2: Example outcome of lesion identification in four spinal cords in a lumbar segment. From left to right the grades are control, low grade, intermediate grade and high grade of EAE. For each of two subplots, the upper image represents a raw $T_2$-map, while the lower image shows the same map with the manual lesion delineation superimposed in yellow.*

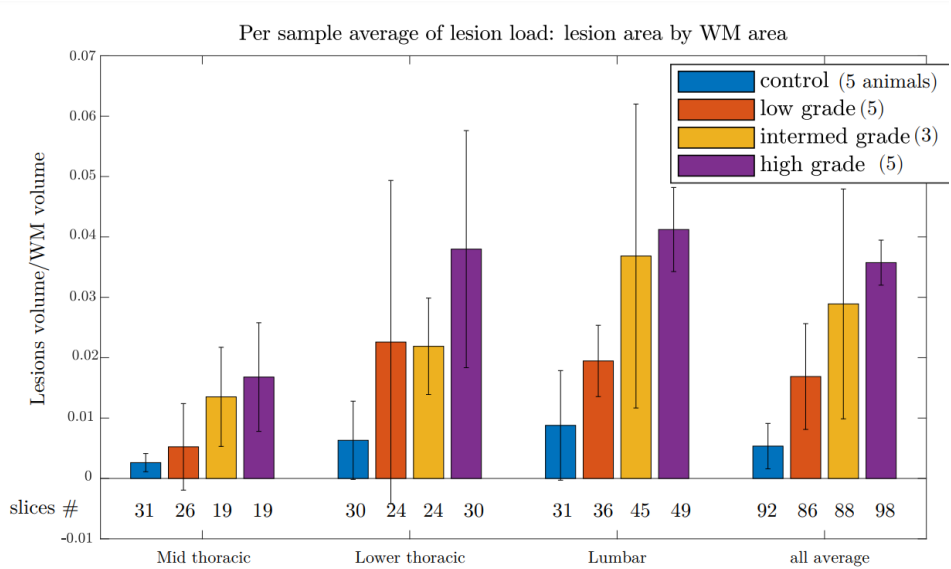

*Figure 3: Correspondence between the grade of animal disability and lesion load in WM of spinal cord tissue. The bar plot shows the lesion load measured by relative volume (number of voxels in lesions divided by number of voxels in the WM of corresponding segment) averaged per sample. Different colors represent distinct grades of EAE. The bar plots from left to right correspond to mid-thoracic, lower-thoracic, lumbar segments of spinal cord. The last barplot represents average over all the segments. Each voxel corresponds to $6.1 \cdot 10^{-4}$ mm$^3$, an average segment volume is ~2000-3000 voxels for mid thoracic, ~4000-5000 for lower thoracic and 6000-8000 for lumbar segments. Each bar plot is an average of 5 samples (low-grade), 3 samples (intermediate grade), 5 samples (high grade), 5 samples (control), 18 samples in total. Error bars depict standard deviation of values within samples. In a row under each one of the bars number of slices in particular segment and belong to a particular disability grade is provided. The legend provides number of animals in each of the groups.*

### f. Parameter estimation

The raw images were denoised using the Marchenko-Pastur PCA method (Veraart et al., 2015) and subsequently corrected for Gibbs ringing artefacts (Kellner et al., 2015) before further analyses. Twenty-two independent components of diffusion and kurtosis tensor (Jensen et al., 2005) were fit to the data using Levenberg-Marquard weighted linear least squares (Veraart et al., 2013). Based on (Chuhutin et al., 2017), WM voxels were fit up to a maximum $b$-value of $b_{\max} = 2.5\,\mathrm{ms\,\mu m^{-2}}$, and GM voxels were fit up to $b_{\max} = 1.2\,\mathrm{ms\,\mu m^{-2}}$. The fit quality was inspected for each sample. An example of data fit for a representative voxel in WM lesion, NAWM and GM is shown in Fig. 1 (C,D). Diffusion and kurtosis tensor parameters were calculated according to (Hansen et al., 2014, 2013; Jensen et al., 2005; Jensen and Helpern, 2010). The exact analytical derivations of WMM parameters from the elements of diffusion and kurtosis tensors used in this study are provided in (Jespersen et al., 2017) (assuming a Watson distribution of neurites). The general case is presented in (Novikov et al., 2018). Different sets of WMM parameters can yield the same DKI signal, an effect known as degeneracy, and a matter of current interest (Jelescu et al., 2015a, 2015b). However, in this work only parameters corresponding to the so-called 'plus' branch (Hansen and Jespersen, 2017; Jespersen et al., 2017; Novikov et al., 2016), typically having $D_a > D_{e,\|}$ were considered. Less than 10% of voxels in any slice displayed non-physical values, such as a negative diffusivity. These voxels were excluded from further statistical analysis of WMM parameters. In total, for all spinal cords, 245851 GM voxels and 246393 WM voxels were analyzed for DTI/DKI parameter estimation, while WMM parameters were estimated in 232274 voxels.

In WM, the estimated parameters were: axial diffusivity ($D_\|$), radial diffusivity ($D_\perp$), fractional anisotropy (FA), axial kurtosis ($K_\|$), radial kurtosis ($K_\perp$), and the previously mentioned WMM parameters (extra-axonal radial $D_{e,\perp}$ and axial $D_{e,\|}$ diffusivities, intra-axonal diffusivity $D_a$, volume fraction of axonal compartment $f$, and concentration parameter of the Watson distribution, $\kappa$). In GM the low tissue anisotropy causes the estimated direction of primary eigenvector to be unstable/poorly defined, and thus, the values of axial and radial diffusivity and kurtosis are less reliable/meaningful. Due to that and in order to restrict the number of compared parameters to avoid unnecessary multiple comparisons, it was decided to limit the scope of estimated parameters in GM to MD and MKT.

### g. Statistical Analysis

The voxels from all spinal cords were input to a linear mixed effects model (LME) (Gelman and Hill, 2007; Goldstein, 2011). The choice of model was guided by current recommendations in (Barr et al., 2013; Bolker et al., 2009) and iterative maximization of Akaike information coefficient (Akaike, 1998). Each of 12 examined parameters $p_i$ was thus fit to

$$p_i \sim g \cdot s + l + (s \cdot g \,|\, a) + (l\,|\,a) \qquad \text{(Equation 3)}$$

using Wilkinson notation (Wilkinson and Rogers, 1973), where $g$ is grade, $s$ is slice, lesion is $l$, $a$ is sample (animal). The 'fixed' effects part of the model was designed to allow the parameters to depend on grade, while the size of the effect was permitted to be different in various SC segments (first term). The second term encodes the expected difference in parameter values inside and outside the $T_2$ hyperintense lesions. Sample-to-sample variations were allowed by including 'random' effects for segment, grade and lesion, each grouped sample-wise.

To avoid a small number of data points having an undue influence on the regression, outliers 2.5 standard deviations above and below the model residual means, were removed after the initial fit, and the model was refitted. This procedure was in agreement with literature (Baayen, 2008; Baayen et al., 2008; Tremblay and Tucker, 2011). The removed outliers attributed for less than 4% of data. The distribution of quality of fit parameters ($\chi^2$) was comparable to the rest of the data.

For each of the 'fixed' effects, analysis of variance (ANOVA) p-values were calculated post hoc. These p-values represent the significance of individual fixed effects as well as the combined effect of segment and grade on parameter. The p-values describing the significance of the linear relationship between the measured parameter and the grade of disability of the EAE animal were finally reevaluated using false discovery rate (FDR) procedure (Benjamini and Hochberg, 1995).

The quality of the fit of LME was estimated using $R^2_\beta$ (Edwards et al., 2008), so that

$$R^2_\beta = \frac{(q-1)\,\nu^{-1} F\left(\hat{\beta}, \hat{\Sigma}\right)}{1 + (q-1)\,\nu^{-1} F\left(\hat{\beta}, \hat{\Sigma}\right)} \qquad \text{(Equation 4)}$$

where $F\left(\hat{\beta}, \hat{\Sigma}\right)$ is a statistic corresponding to the null hypothesis $H_0: \beta_1 = \beta_2 = \ldots = \beta_{q-1} = 0$ for $q-1$ fixed effects $\beta_i$, $\nu$ is Satterthwaithe estimator of degrees of freedom. Partial $R^2_\beta$ were calculated to obtain the relative measure for each of the 'fixed' effects with a statistics corresponding to the null hypothesis $H_0: \beta_j = 0$ for $j \in \{1, \ldots, q-1\}$.

For the post-hoc analysis, the means of the parameters were calculated for each sample for each SC segment, and for the SC as a whole, in GM and WM separately. One-way ANOVA was used to evaluate the significance of difference between the grades. The parameters surviving FDR correction were checked and the grades with significantly different means were identified using post-hoc two sample tests.

A further post-hoc comparison of individual parameters inside lesions supported the initial LME model assumption of independence between the grade and lesion LME model parameters (data not provided). Thus, we based the post-hoc lesion analysis on the premise that the distribution of parameters inside lesions does not depend on segment or EAE grade.

## IV. Results

### a. Clinico-radiological paradox: Grade and lesion load

Figure 3 shows the relationship between the lesion load in each segment and EAE grade. A small number of voxels were marked as lesion in control animals. This did not differ substantially between different segments (group-wise one-way ANOVA), and was most likely due to human classification error.

The low-grade spinal cords exhibited a visible difference in lesion load between medium thoracic segments compared to lower thoracic and lumbar segments, where on average 5 times as many voxels were affected by lesions. There was a slightly lower lesion load in lumbar than in lower thoracic segment for this grade. Animals with intermediate and high grade of disability shared the same pattern of increase in lesion load in the caudal direction; however, a high variance in the measurements prevented this effect from being statistically significant.

A one-way-ANOVA of sample-wise mean of relative lesion load showed that controls were significantly different from the diseased (EAE) animals in all segments. However, the difference between grades of disability was not statistically significant.

### b. Diffusion MRI: Parameter estimation

Figure 4 shows parameter maps of all the investigated parameters for a representative animal in each of the grades (control, low, intermediate, and high grade) in the medium thoracic (T9) segment. WMM parameters are restricted to the manually delineated WM to approximately fulfill the assumptions of the model. Qualitatively, the maps show an increase in asymmetry in animals with higher disability grade. Most of the maps provide sufficient contrast between the lesions. The Watson concentration parameter $\kappa$ displays the biggest variation in the maps.

Quantitative results (mean and standard deviation along the disability group) for all the measured parameters are provided in Tab. 1.

### c. Diffusion MRI: validating metrics with LME

Table 2 shows the estimators of the LME fit quality and values that quantify the capability of LME model parameters to explain each of the parameters measured voxel-wise.

All parameters demonstrated relatively good quality of fit $R^2_\beta$. MD and $D_{e,\parallel}$ attained the lowest values of ~0.91.

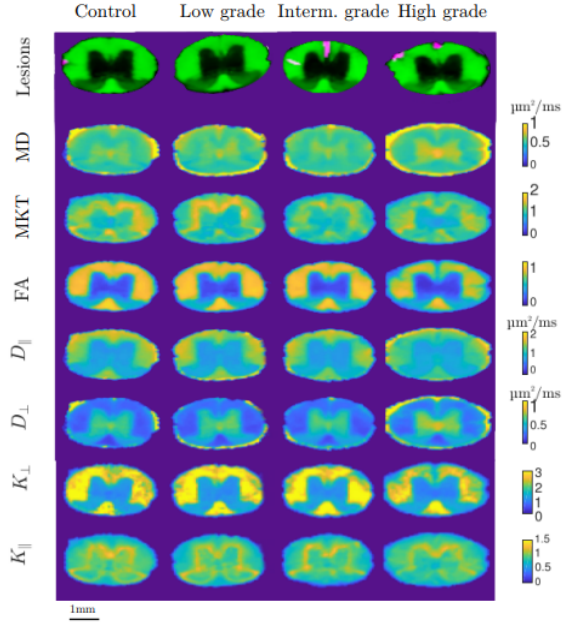
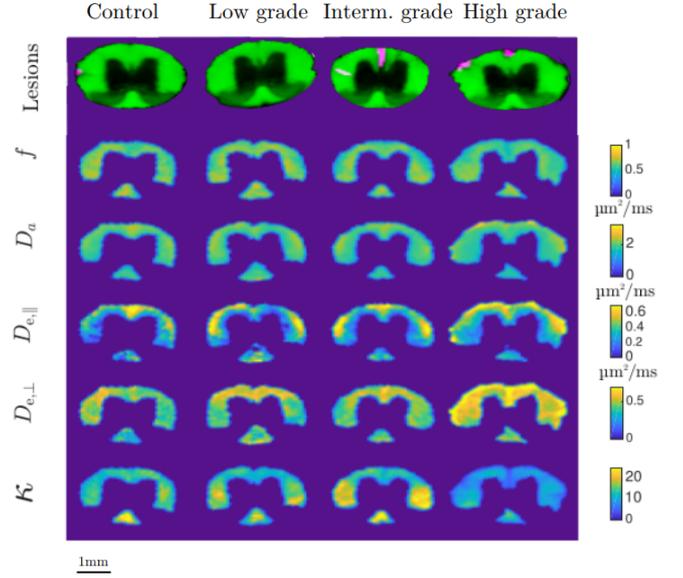

Figure 4: Examples of parameter maps for each of the measured parameters in mid-thoracic segments of spinal cord. Each column (from left to right) corresponds to different grades of EAE disability: control animal, low grade, intermediate grade and high grade of EAE. Rows correspond to different measured parameters (A) (from top to bottom): mean diffusivity, MKT, FA, axial diffusivity, radial diffusivity, radial kurtosis, parallel kurtosis; (B): axonal water fraction, axonal diffusivity, axial extra-axonal diffusivity, radial extra-axonal diffusivity and concentration parameter of Watson distribution, the upper row depicts the delineation of spinal cord on the background of FA map.

| Tissue | Parameter | Control | | | | Low grade | | | | Intermed grade | | | | High grade | | | |
|---|---|---|---|---|---|---|---|---|---|---|---|---|---|---|---|---|---|
| | | NAWM | | Lesion | | NAWM | | Lesion | | NAWM | | Lesion | | NAWM | | Lesion | |
| | | Value | Stdev | Value | Stdev | Value | Stdev | Value | Stdev | Value | Stdev | Value | Stdev | Value | Stdev | Value | Stdev |
| GM | MD | 0.57 | 0.06 | | | 0.58 | 0.06 | | | 0.58 | 0.06 | | | 0.58 | 0.07 | | |
| GM | MKT | 1.15 | 0.18 | | | 1.17 | 0.18 | | | 1.15 | 0.18 | | | 1.10 | 0.20 | | |
| WM | $K_\perp$ | 2.66 | 0.71 | 1.89 | 0.52 | 2.57 | 0.65 | 1.44 | 0.46 | 2.36 | 0.64 | 1.37 | 0.37 | 2.20 | 0.67 | 1.37 | 0.43 |
| WM | FA | 0.80 | 0.06 | 0.48 | 0.17 | 0.79 | 0.07 | 0.35 | 0.15 | 0.74 | 0.10 | 0.35 | 0.11 | 0.74 | 0.11 | 0.40 | 0.14 |
| WM | $K_\parallel$ | 0.91 | 0.14 | 1.04 | 0.17 | 0.87 | 0.13 | 0.96 | 0.14 | 0.89 | 0.14 | 0.89 | 0.12 | 0.87 | 0.15 | 0.94 | 0.12 |
| WM | $D_\perp$ | 0.21 | 0.05 | 0.47 | 0.20 | 0.23 | 0.06 | 0.52 | 0.17 | 0.26 | 0.08 | 0.52 | 0.12 | 0.26 | 0.08 | 0.45 | 0.10 |
| WM | $D_\parallel$ | 1.24 | 0.16 | 1.14 | 0.20 | 1.26 | 0.16 | 0.98 | 0.21 | 1.21 | 0.18 | 0.97 | 0.16 | 1.19 | 0.20 | 0.91 | 0.17 |
| WM | $f$ | 0.64 | 0.07 | 0.54 | 0.08 | 0.64 | 0.06 | 0.51 | 0.09 | 0.61 | 0.06 | 0.49 | 0.07 | 0.61 | 0.07 | 0.48 | 0.08 |
| WM | $D_a$ | 1.94 | 0.22 | 1.93 | 0.34 | 1.95 | 0.21 | 1.98 | 0.38 | 1.90 | 0.21 | 1.98 | 0.27 | 1.91 | 0.24 | 1.86 | 0.26 |
| WM | $D_{e,\parallel}$ | 0.36 | 0.19 | 0.45 | 0.18 | 0.38 | 0.19 | 0.50 | 0.25 | 0.39 | 0.16 | 0.55 | 0.18 | 0.41 | 0.19 | 0.49 | 0.17 |
| WM | $D_{e,\perp}$ | 0.48 | 0.12 | 0.66 | 0.22 | 0.49 | 0.11 | 0.78 | 0.23 | 0.54 | 0.12 | 0.74 | 0.16 | 0.49 | 0.13 | 0.64 | 0.15 |
| WM | $\kappa$ | 12.54 | 3.92 | 5.31 | 2.58 | 12.67 | 4.17 | 3.83 | 2.04 | 9.90 | 3.88 | 3.52 | 1.83 | 10.82 | 4.55 | 4.08 | 1.96 |

Table 1: Fit results of all measured parameters averaged for disability group. For each parameter a mean estimate provided along with standard deviation of error inside and outside the lesion (in NAWM). The mean and standard deviation of error was calculated only in the tissues in which a particular parameter was used for a successive LME analysis. Thus the statistics for MKT and MD was estimated only in GM, and the statistics for the rest of the DKI and WMM parameters was estimated only in WM. Note also that GM parameters (MD, MKT) were not calculated inside the lesions.

P-values shown in Tab. 2 quantify the extent to which each of the 12 studied parameters can be explained by the parameters of the linear mixed effects model. Grade had a significant effect on 7 out of 12 parameters after FDR: MKT in GM, and $K_\perp$, $D_\perp$, $D_{e,\parallel}$, $D_{e,\perp}$, $\kappa$ and $f$ in WM. All parameters except MD and MKT in GM and $f$ in WM were found to depend significantly on the segment. Likewise, the interaction between grade and segment was statistically significant in all but five parameters, i.e. the two GM parameters MD and MKT, three WM parameters $D_{e,\parallel}$, $D_a$ and $D_\parallel$. All parameters but $D_a$ were significantly different between lesion and normal appearing brain tissue.

The results of the calculation of partial $R_\beta^2$ (Edwards et al., 2008) for each fixed effect variable are also provided in Tab. 2. $R_\beta^2$ revealed high association between the kurtosis/WMM parameters and the disability grade of EAE (~0.9) for all the parameters that were found significant in FDR procedure. The comparison of partial $R_\beta^2$ values showed that the disability grade accounted for most of the variation in 4 out of 12 parameters $D_\perp$, $D_{e,\parallel}$, $D_{e,\perp}$ in WM and MKT in GM.

Additional characteristics of the LME fit are provided as Supplementary material. These include a different measure of fit quality (Johnson, 2014; Nakagawa et al., 2013) and estimates and confidence intervals of fixed effects of LME-model. These estimates show that among the parameters which are significantly correlated with the grade MKT, $K_\perp$, $f$, $D_{e,\perp}$ and $\kappa$ decrease with the increase in disability grade, while $D_\perp$ and $D_{e,\parallel}$ increase with increasing grade.

### d. Diffusion MRI: Post-hoc statistical analysis

From the LME analysis, we found that the variation of several GM and WM parameters can be explained by EAE-grade and by lesion status, i.e. whether or not the voxel is located inside a lesion. A follow-up post-hoc analysis intended to investigate group-wise behavior of the segment-wise means in parameters with a significant grade. In particular, Table 3 shows the results of the post-hoc analysis of sample means outside the lesions, in GM and NAWM.

In GM, MKT showed significant difference between the grades only in the slices located in mid-thoracic segment of the spinal cord, specifically between low and intermediate and between low and high grades.

In NAWM, 5 out of the 6 biomarkers surviving FDR correction demonstrated significant differences between the control and diseased animals, mainly in the lumbar SC. Two DKI parameters ($K_\perp$, $D_\perp$), and two WMM parameters ($f$, $D_{e,\parallel}$) depended significantly on EAE grade.

| Tissue | name | Outliers % | $R^2_\beta$ | P-values | | | | partial $R^2_\beta$ | | | | Grade FDR |
|---|---|---|---|---|---|---|---|---|---|---|---|---|
| | | | | Grade | Lesion | Segment | Segment:Grade | Grade | Lesion | Segment | Segment:Grade | |
| GM | MD | 1.7 | 0.910 | 0.576 | ✕ | 0.060 | 0.382 | 0.117 | ✕ | 0.588 | 0.295 | |
| GM | MKT | 1.8 | 0.980 | 0.010 | ✕ | 0.136 | 0.290 | 0.863 | ✕ | 0.711 | 0.663 | * |
| WM | $K_\perp$ | 2.1 | 0.995 | <0.001 | <0.001 | 0.011 | 0.004 | 0.953 | 0.991 | 0.896 | 0.839 | * |
| WM | FA | 3.5 | 0.977 | 0.042 | <0.001 | 0.002 | 0.004 | 0.852 | 0.982 | 0.972 | 0.852 | |
| WM | $K_\parallel$ | 1.9 | 0.979 | 0.050 | 0.001 | 0.001 | 0.016 | 0.678 | 0.864 | 0.922 | 0.830 | |
| WM | $D_\perp$ | 2.1 | 0.992 | <0.001 | 0.000 | 0.003 | 0.021 | 0.972 | 0.926 | 0.955 | 0.765 | * |
| WM | $D_\parallel$ | 1.7 | 0.991 | 0.590 | 0.000 | 0.003 | 0.378 | 0.132 | 0.903 | 0.838 | 0.408 | |
| WM | $f$ | 2.2 | 0.962 | <0.001 | <0.001 | 0.582 | 0.011 | 0.967 | 0.971 | 0.259 | 0.853 | * |
| WM | $D_a$ | 1.5 | 0.971 | 0.525 | 0.123 | 0.013 | 0.110 | 0.204 | 0.478 | 0.862 | 0.543 | |
| WM | $D_{e,\parallel}$ | 1.4 | 0.919 | 0.003 | 0.014 | 0.004 | 0.349 | 0.886 | 0.738 | 0.923 | 0.412 | * |
| WM | $D_{e,\perp}$ | 2.4 | 0.948 | 0.008 | <0.001 | 0.003 | 0.005 | 0.949 | 0.897 | 0.882 | 0.910 | * |
| WM | $\kappa$ | 2.1 | 0.993 | <0.001 | <0.001 | <0.001 | <0.001 | 0.957 | 0.992 | 0.980 | 0.916 | * |

Table 2: The results of the fit of the linear mixed effects model. For each of the studied parameters (in rows), the following are presented in columns: percent of outlier values removed, quality of LME fit $R^2_\beta$ (Edwards et al., 2008), p-values for coefficients of grade, lesion, segment and segment*lesion, partial $R^2_\beta$ (Edwards et al., 2008) of the same four coefficients and the results of the FDR multiple comparison test. Since lesions were registered only in WM, the coefficients of lesion are absent in GM.

### Table 3 (A)

| | Low grade | Intermediate grade | High grade |
|---|---|---|---|
| Low grade | | MTO: <u>MKT</u> | MTO: <u>MKT</u> $K_\perp$ $f$ $D_{e,\parallel}$<br>LTO: $K_\perp$ $f$<br>LU: $K_\perp$ $D_\perp$ $f$<br>All: $K_\perp$ $f$ |
| Intermediate grade | | | MTO: $K_\perp$ |

### Table 3 (B)

| | Low grade | Intermediate grade | High grade |
|---|---|---|---|
| Control | MTO:<br>LTO:<br>LU:<br>All: | MTO:<br>LTO:<br>LU: FA $K_\perp$ $D_\perp$ $f$<br>All: $D_{e,\parallel}$ | MTO: <u>MKT</u> $K_\perp$ $f$ $D_{e,\parallel}$<br>LTO: $K_\perp$ $f$ $D_{e,\parallel}$<br>LU: FA $K_\perp$ $D_\perp$ $f$ $D_{e,\parallel}$ $\kappa$<br>All: $K_\perp$ $D_\perp$ $f$ $D_{e,\parallel}$ |

Table 3: Post-hoc analysis of parameters in GM and NAWM. Average value for all the NAWM or GM in the particular segment in each sample. (A) For each of the disability groups comparisons low-grade vs intermediate grade, low-grade vs high grade and intermediate grade vs high grade parameters that were found significant (p<0.05) after ANOVA of per-sample mean in each one of the segments is provided in the corresponding cell. An FDR correction of multiple comparisons has been taken into account. GM parameters are underlined. (B) Each of disability groups (low, intermediate, high) compared with the control group. Parameters that were found significant (p<0.05) after ANOVA of per-sample mean in each of the segments are listed in corresponding cells. An FDR correction has been performed. GM parameters are underlined.

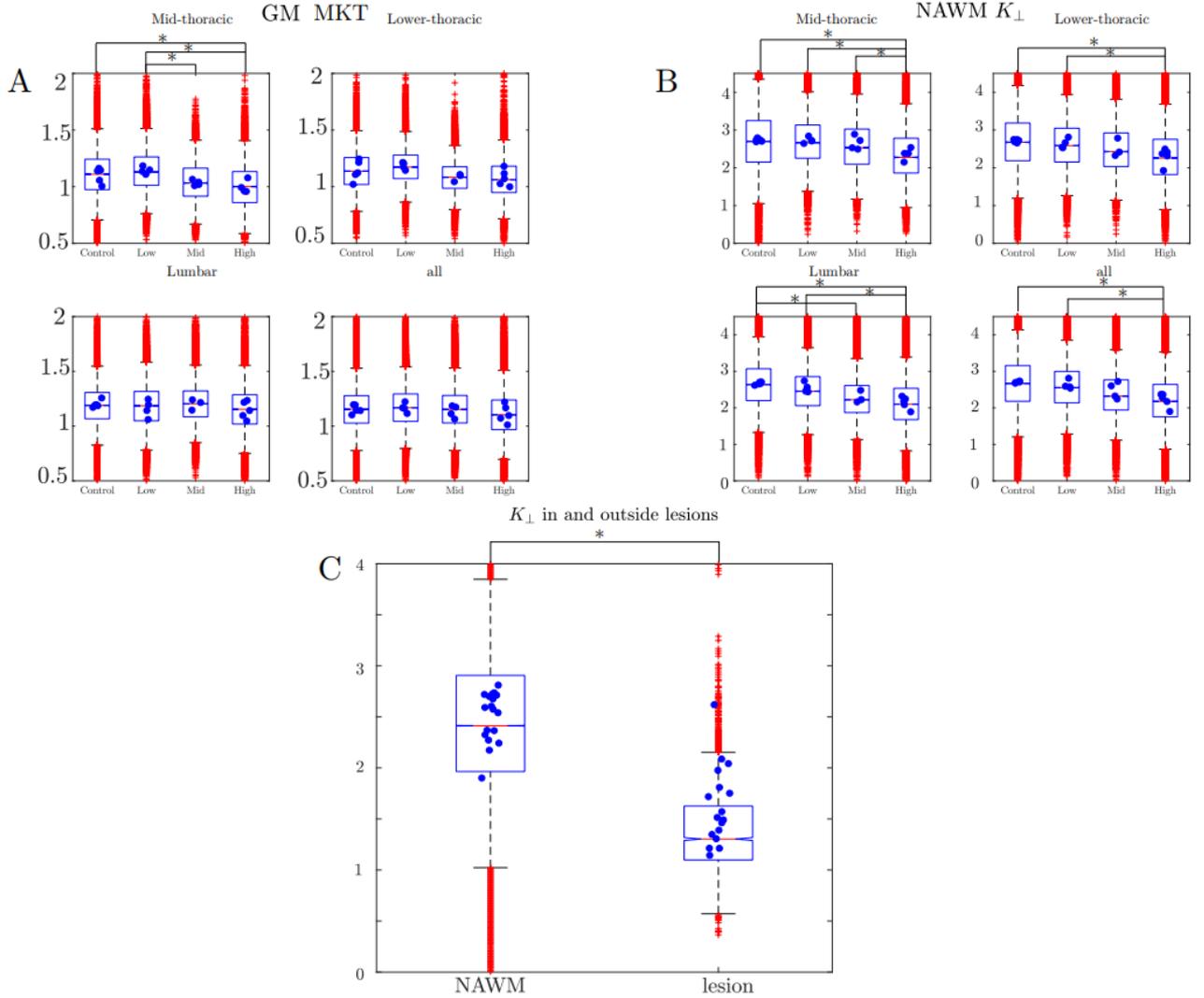

Figure 5: Examples of parameter distributions in the post-hoc parameter analysis for MKT in GM (A) and $K_\perp$ in NAWM (B) illustrated with box-plots. Each of four subplots corresponds to one of the spinal cord segments (mid-thoracic, low-thoracic and lumbar and a graph for all voxels pooled across segments). Each box-plot represents parameter distribution for a corresponding EAE-grade (control animals, low-grade, medium grade, high grade). Blue dots correspond to the parameter means within each spinal cord. Asterisk denotes significant group-wise difference between the spinal cord means, tested with ANOVA as described in post-hoc analysis (C) illustrates the difference between the NAWM and lesion tissue in $K_\perp$. Each box-plot represents the distribution of values inside and outside the hyperintensity lesions. Blue dots correspond to the parameter means within each spinal cord. Asterisk denotes significant group-wise difference between the spinal cord means. In all three plots the central mark indicates the data median, the bottom and top edges indicate the 25th and 75th percentiles. The whiskers extend to the most extreme data points excluding outliers, and the outliers (voxels) are plotted individually in red.

$K_\perp$ and $f$ were found to survive pooling all segments together, demonstrating an overall significant difference between high and low grade.

As an illustration, a representative part of the data and the associated post-hoc analysis is given in Fig. 5. In this figure, the distributions over voxels of MKT in GM (Fig. 5 (A)) and $K_\perp$ in NAWM (Fig. 5 (B)) are visualized using box plots for each of the grades and segments. The means of the spinal cords, which were used in the post hoc analysis (Tab. 3), are superimposed on the boxplots as blue circles. The significantly different grades are marked by asterisks. Note that the seemingly large number of outliers apparent in the boxplots of Fig. 5 constitute a small fraction of the more than 10000 voxels sampled for each spinal cord.

The same type of post-hoc analysis that was used to study the voxels outside of the lesions, was used to investigate voxels inside the lesions. The analysis revealed that the vast majority of segment-wise means inside the hyperintensity lesions did not show any significant differences between EAE-grades, with only $K_\parallel$ in lower thoracic segment showing difference between the grades at the level of significance $p < 0.05$ (results of this analysis are provided as Supplementary material).

Since the LME analysis revealed that lesion status had a significant effect on most of the estimated parameters, a post-hoc analysis was adapted to test for difference in parameter means between lesions and NAWM. The results of this comparison are provided in Tab. 4. The difference between the sample-wise means of NAWM and lesions voxels was found to be significant for all 10 WM biomarkers. No difference was found between various segments.

Figure 5 (C) illustrates the difference between NAWM and lesion tissue in $K_\perp$. The mean in each spinal cord is plotted with blue dots.

### e. Combining T$_2$ and diffusion MRI

Based on the previous results, we next consider a "hybrid" way of addressing the clinico-radiological paradox, i.e. using a compound variable that reflects both lesion load and NAWM health to provide a way of distinguishing different grades of EAE. In particular, an animal-wise LME fit of the model

$$g = p_0 + p_1 \cdot l + p_2 \cdot \hat{K}_\perp + p_3 \cdot \hat{D}_{e,\perp} \qquad \text{(Equation 5)}$$

where $g$ is the grade and $l$ is lesion load, $\hat{D}_{e,\parallel}$ and $\hat{K}_\perp$ are animal-wise mean values of $D_{e,\parallel}$ and $K_\perp$ in NAWM, yielded coefficient values $p_0 = 2.9, p_1 = 12.7, p_2 = -2.3, p_4 = 10.4$. A "hybrid" metric that uses these parameters is able to distinguish not just between control and high, control and intermediate groups but also between low and high, low and intermediate EAE-grades. The results of using such metric are shown in Fig. 6. However, a follow up study that will test this hybrid metric with an independent data is needed.

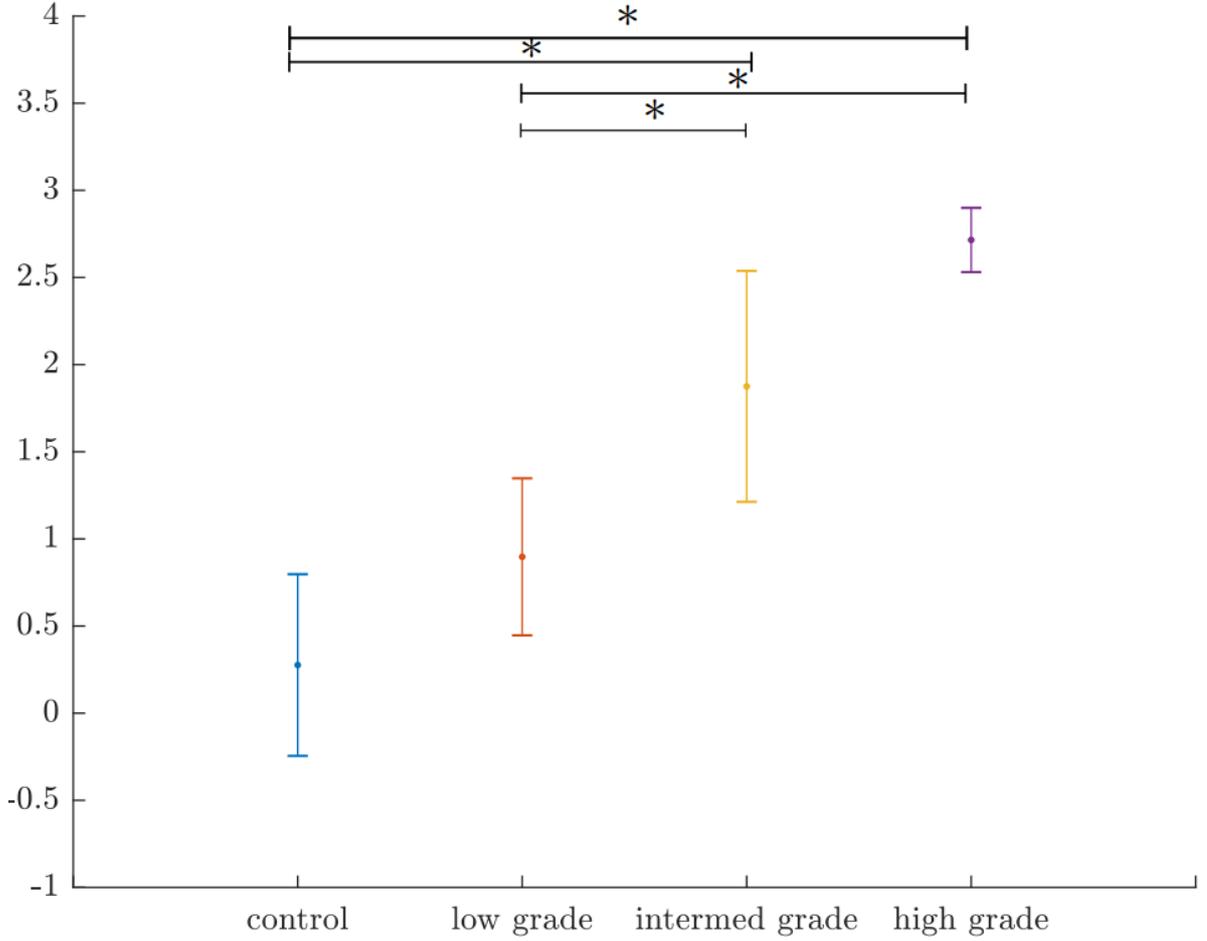

Figure 6: Application of a "hybrid" biomarker (Eq. 6) on animal-wise data. From left to right the values are group-wise means of described control, low-grade, intermediate and high grade EAE. Error bars depict standard deviation of animal-wise estimates of the biomarker. Asterisk denotes statistical significance measured with 1-way ANOVA.

| FA | $K_\parallel$ | $K_\perp$ | $D_\parallel$ | $D_\perp$ | $f$ | $D_\mathrm{a}$ | $D_{e,\parallel}$ | $D_{e,\perp}$ | $\kappa$ |
|---|---|---|---|---|---|---|---|---|---|
| <0.001 | 0.001 | <0.001 | <0.001 | <0.001 | <0.001 | 0.0497 | 0.012 | <0.001 | <0.001 |

Table 4: Post-hoc analysis of NAWM against lesion tissue. For each WM parameter (in columns) we present the p-values that characterize the significance of difference in per-sample parameter mean inside against outside the lesion.

# V. Discussion

Mapping quantitative biomarkers for MS – as well as other neurological disorders – is required for early diagnosis, follow up of treatments, and testing novel avenues for treating disease. In this context, MRI is highly beneficial due to its clinical safety and versatility. In the context of MS, lesion load and tissue atrophy are quite easily imaged both in humans and in preclinical studies in animals and tissues. However, atrophy is typically a very late marker, and, even though lesion load has been historically associated with motor deficits in MS and EAE (Bjartmar et al., 2001; Sathornsumetee et al., 2000), its correlation with disability is poor (Barkhof, 2002; Bergers et al., 2002; Robinson et al., 2010; Wuerfel et al., 2007), a disparity known as the clinico-radiological paradox (Cohen et al., 2016; Nathoo et al., 2014). Diffusion-based metrics are promising as they map microstructural aspects of the tissue. Therefore, this study sought to test quantitative metrics obtained from DKI against disease severity in an animal model of MS.

Consistent with the clinico-radiological paradox, we did not find correlations between the EAE-grades and lesion load, although the lesion load in the EAE animals was significantly different to that in the control group. This is perhaps not surprising given the similarities shared between the EAE model and actual MS. We therefore turned to LME fitting of the more comprehensive DKI data.

The choice of LME to estimate and study the effects of EAE on kurtosis tensor parameters was shaped by the observations that pathological changes both inside lesions, in NAWM and in GM contribute to clinical disability in both human MS and in animal models of neurodegeneration (de Kouchkovsky et al., 2016; Evangelou et al., 2000; Filippi et al., 2012; Filippi and Rocca, 2011; Inglese and Bester, 2010; Kipp et al., 2016; Lassmann and Bradl, 2016). The specific form of the LME designed to take into account random contributions of sample to sample variability. Our assessment of LME fitting quality (in Tab. 2) was in line with up-to-date recommendations for LME (Baayen et al., 2008; Edwards et al., 2008; Nakagawa et al., 2013).

In GM, MKT depended significantly on disability grade (Tabs. 2,3; Fig. 5). This is in line with human studies (Agosta et al., 2007; Bester et al., 2015; Raz et al., 2013; Zackowski et al., 2009) reporting similar changes in GM. Such changes are likely indications of GM pathology, possibly associated with neuronal degeneration and myelin loss in neurites (Guglielmetti et al., 2016), and reinforce the involvement of GM in EAE disability. Interestingly, while the biggest burden of lesions (Fig. 2) and most of the changes in NAWM (Tab. 3; Fig. 5) were associated with the lumbar section of the spinal cord, most of the changes detected in GM were observed in mid-thoracic sections. Given that the EAE-induced disability progresses from hind- to forelimbs, one might hypothesize that the damaged GM tissue in mid thoracic SC is connected to the damaged fascicles in lumbar WM. Thus, in EAE, correlated pathological mechanisms may be responsible for damage in NAWM and GM. This may be similar to human MS where the spatial and temporal relationships between the damage in GM and NAWM are still not fully resolved and might depend on disease phenotype (Bodini et al., 2009; Pirko et al., 2007; Steenwijk et al., 2015; Tewarie et al., 2018). Future longitudinal studies could shed light onto this mechanism and elucidate whether GM damage is a precursor of future damage in NAWM.

In NAWM, $K_\perp$ and $D_\perp$ showed the strongest and most robust results among tissue biomarkers derived from kurtosis and diffusion tensors. Radial kurtosis showed the strongest inverse relationship with EAE grade (i.e. it decreased with increasing disease severity). Such changes have also been observed previously in preclinical models of MS (Falangola et al., 2014; Jelescu et al., 2016; Kelm et al., 2016), while the opposite effect was observed in (Guglielmetti et al., 2016). Our result might suggest closer similarities of the EAE to cuprizone or genetically induced chronic demyelination than to the acute inflammatory demyelination used in (Guglielmetti et al., 2016). An increase in $D_\perp$ was also found to be significantly correlated with EAE grade. Again, this behavior agrees with previous chronic demyelination studies (Falangola et al., 2014; Jelescu et al., 2016; Kelm et al., 2016) and with results of a previous DTI EAE study (Budde et al., 2009). Early human studies demonstrated similar behavior of $D_\perp$ and associated it with demyelination (Klawiter et al., 2011) and possible axonal loss (Naismith et al., 2010).

Among WM model parameters, $D_{e,\|}$ was the one affected the most by EAE grade, while $D_{e,\perp}$ was affected in a much weaker manner and with no significant effects in post-hoc analysis (Tabs. 2,3). Counter-intuitively, an increase in $D_{e,\|}$ with grade was found. This fact could potentially be explained by axonal damage, changes in the structure of glial cells, and myelin loss, causing the extra-axonal space to have lower tortuosity and consequently a higher diffusivity. This result is in contrast with cuprizone models (Falangola et al., 2014; Guglielmetti et al., 2016; Jelescu et al., 2015b). The disparity can stem from differences between the mechanisms underlying tissue degeneration but also from other microstructural differences between the neural tissue in cerebrum and in spinal cord. Alternatively, it may also be a result of choosing a different solution 'branch' when finding parameters of WMM model (Hansen et al., 2017; Jelescu et al., 2015b; Jespersen et al., 2017).

The axonal water fraction was also significantly affected by the differences in EAE grade. This parameter ($f$) has been suggested to be a biomarker of axonal loss (Fieremans et al., 2011). We found $f$ to decrease with an increase in EAE-grade, therefore an axonal loss in NAWM could be one driver of the disability. The post-hoc results showed that along with $K_\perp$, changes in the mean values of $f$ were distributed homogeneously among all the segments of spinal cords; consequently axonal loss could be a diffuse feature in thoracic and lumbar SC.

The ratio $\lambda = D_{e,\parallel}/D_{e,\perp}$ (tortuosity) has been proposed as a biomarker of demyelination (Fieremans et al., 2012). In our data (see Results and Supplementary material), $D_{e,\parallel}$ increased strongly and $D_{e,\perp}$ decreased, thus overall tortuosity increased with increase in grade (the effect of grade on tortuosity variation was found to be significant in post-hoc LME fit, data available upon request). This finding is in contrast to (de Kouchkovsky et al., 2016; Falangola et al., 2014) and may provide evidence of pathological processes in EAE.

Our data shows that the Watson concentration parameter $\kappa$ significantly decreased in a way that could be explained by EAE grade. This might be a result of axonal damage that could cause the breaking of fascicles and fanning out of individual axons. According to post-hoc analysis, this behavior was present in the lumbar segment of the spinal cord. Our result is in line with (Schneider et al., 2017) where increased fiber dispersion in NAWM of MS SC was reported. However, increased fiber dispersion is also present inside the lesions in mouse SC. This is in contrast with (Grussu et al., 2017), where a decrease in neurite orientation dispersion was measured in lesions of MS using neurite orientation dispersion and density imaging (NODDI) and histology. One possible reason for the disparity is different species (animal and human), where different pathological mechanisms could be at play. Another reason could be related to hypomyelinating lesions being not as well defined in rodent models, and in EAE in particular, as in humans. Therefore, even though lesion detection was performed in a consistent and 'blind' way, a systematic error may have been introduced, e.g. if too big portions of NAWM are segmented as lesions. Third, differences in the employed diffusion models could be responsible for the disparity. Validation of fiber dispersion using microscopy is needed in order to address this discrepancy.

$D_\parallel$ has previously been shown to decrease significantly with EAE score and with axonal injury (Budde et al., 2009). Both $K_\parallel$ and $D_\parallel$ were significantly affected in some cuprizone studies (Falangola et al., 2014; Guglielmetti et al., 2016), but not in (Jelescu et al., 2016; Kelm et al., 2016). Our study found no evidence of any correlation between EAE grade and $D_\parallel$ or $K_\parallel$. Consequently, our work provides an indication that in EAE, tissue changes due to demyelination and axonal loss are insufficient to change diffusivity or kurtosis parallel to fiber bundles.

In our study, FA showed no significant dependence on grade. This observation is in line with (Guglielmetti et al., 2016) where FA was not able to differentiate between the treatment groups and control. Based on our results, FA can be well explained by lesion load and therefore may also be a good biomarker of lesion burden, however it has a low utility in NAWM.

Both the results of LME model fitting and post-hoc analysis demonstrated that all parameters but one ($D_a$) were able to distinguish lesions from NAWM. However, the value of lesion detection using model parameters may be limited, due to the clinico-radiological paradox as formulated in (Nathoo et al., 2014). In addition, lesion delineation using $T_2$ maps is far more convenient due to faster and easier acquisition protocols. An interesting finding of this work is that lesions do not differ across grades, since the vast majority of the parameters within lesions did not differ. A probable explanation of that is that CNS tissue that makes up WM lesions does not contribute to disability in EAE.

Recent studies (By et al., 2018, 2017; Grussu et al., 2015; Schneider et al., 2017) have applied NODDI and spherical means (SMT) techniques to spinal cord tissue WM in healthy controls and in MS patients and demonstrated promising diagnostic results. However, our findings suggest that the basic assumptions of these studies could be violated. In particular, since $D_a$ was found not to be driven by EAE grade while $D_{e,\|}$ increased with grade, those two parameters cannot be in constant as assumed in NODDI ( $D_a = D_{e,\|}$ =1.7 $\mu m^2\,ms^{-1}$). The constant tortuosity assumption (imposed in both NODDI and SMT) was also not valid in our dataset. Thus, assuming that our results are valid in human MS, the values estimated with those two techniques would be unable to reveal the true microstructural changes associated with disease progression and disability. Releasing all the parameters of the so-called 'standard' model (Novikov et al., 2018; Jelescu et al., 2015a; Jespersen et al., 2007) may thus prove necessary.

The parameters of WMM observed here were different to prior publications. This is perhaps not surprising given differences in the studied tissues and the disease inception mechanisms. With the current rate of improvement in MRI techniques, parameters of the full WMM in human MS spinal cord can soon be estimated and compared to our results to help reassess models of spinal cord pathology in MS. Since the spinal cord and not cerebrum damage is better correlated with accrual of long-term disability (Inglese and Bester, 2010; Lin et al., 2006) the results of such an assessment will improve the understanding of mechanisms of MS progression.

## VI. Limitations

In this work, fixed tissue was used, which allowed longer scanning and better data quality in comparison with in vivo protocols. This choice was justified by the assumption that despite known impact of fixation on tissue properties (Shepherd et al., 2009, 2005; Sun et al., 2005) the pathological effect on damaged tissue will be strong enough to be detectable in present exploration study. Nonetheless, this study suffers from differences between ex-vivo and in-vivo tissues, which have not been fully accounted for yet (Horowitz et al., 2015).

Manual lesion segmentation, even though it was performed 'blindly', can potentially result in a systematic bias in contrast between estimated NAWM and lesion values, as well as increased variability.

Concerns have been expressed (Lampinen et al., 2018) about applicability of modeling constraints required for compartment-based modeling of neural tissue. The compartment models for diffusion used in this work have not been fully validated across different tissue types, in vivo and ex vivo datasets, etc. Thus, the modeling efforts in this study were restricted to ex-vivo mouse spinal cord, where, according to histology (Ong et al., 2008) the axonal size is around 1 μm and in the chosen regime of gradient strengths and waveforms the attenuation due to diffusion along the diameter is negligible (Dyrby et al., 2013; Nilsson et al., 2017). Diffusion times were chosen to be short enough (10ms), so that the exchange has only a minimal effect (Nilsson et al., 2013, 2009). Therefore, there is good reason to believe that in this case, the appraisal of axons as sticks is approximately valid. Further investigation of the validity of the attained results e.g. the role of myelin water and related compartment-dependent $T_2$-relaxation of diffusion weighted signal is necessary before translating the results of this study to the clinic. Data description parameters yielded by kurtosis and diffusion tensor fits were also included for a model independent assessment.

## VII. Clinical implications

This work shows that the disability in EAE and therefore probably also disability in MS is correlated with and maybe is driven by the neural matter outside the lesions.

In line with other works this study also suggests the prominent role of SC GM (Agosta et al., 2007; Bester et al., 2015; Guglielmetti et al., 2016; Raz et al., 2013; Zackowski et al., 2009) and NAWM (Falangola et al., 2014; Jelescu et al., 2016; Kelm et al., 2016) in the development of disability. All these results support the fact that the search for disability biomarkers the human SC should concentrate on the neural matter outside the demyelination lesions.

This study points on the perspective of using neurite tissue models, where extracellular parallel diffusivity and axonal water fraction are recommended for assessment of human MS disability using SC MRI.

There are some barriers in translation of the results of this work to a clinical setting. In addition to human MS pathology being distinct from the EAE animal model in terms of illness outset and its evolution, human scanners feature multiple technical differences compared to the system used here. Such differences can hinder the adaptation of the described methods.

This study was performed with 11 *b*-values and 30 directions, yielding approximately 350 images. However, such a big number of *b*-values was primarily needed to estimate parameters both WM and GM. If a similar approach translated to human studies would focus e.g. only the WM, the same type of analysis can be performed with 5 *b*-values and 30 directions, which can be achieved within a clinically feasible scan time of around 15 minutes. At the same time due to relatively low *b*-values, this protocol is more accessible for clinical systems than two compartment models such as CHARMED (Assaf et al., 2004; Assaf and Basser, 2005; Barazany et al., 2009; De Santis et al., 2016) developed to attempt axonal diameter mapping.

## VIII. Conclusions

- This work has shown that statistical analysis based on linear mixed effect models is capable of disentangling NAWM and lesion effects.
- In $T_2$-hyperintensity WM lesions, none of the measured biomarkers was found to be significantly correlated with EAE-disability.
- In NAWM and GM the relationship between the disability and DKI and DTI metrics was found to be similar to other hypomyelinating MS models and to ex-vivo MS tissue.
- In NAWM, changes in WM-modeling parameters (strong increase in $D_{e,\|}$, weak effect in $D_a$, $D_{e,\perp}$) were clearly different to what has been observed in other animal models of MS.
- This work suggests a potential relationship between the damage in GM and NAWM of EAE SC.
- This work did not detect any significant effect of lesions on EAE-grades, neither using accumulated lesion load nor with DWI biomarkers in the tissue restricted by $T_2$-weighted lesion.
- A strong increase in $D_{e,\|}$ of NAWM is an effect that has not been previously observed in other models of MS.

## IX. Acknowledgements

The authors are grateful for financial support of this project by Lundbeck Foundation Grant R83-A7548 and Simon Fougner Hartmanns Familiefond. AC and BH acknowledge support from NIH1R01EB012874-01.

The authors thank Dr Kevin D Harkins and Prof. Mark D Does from Vanderbilt University for the REMMI pulse sequence and reconstruction toolbox used in this study, which were supported through grant number NIH EB019980. NS was supported in part by the European Research Council (ERC) under the European Union's Horizon 2020 research and innovation programme (grant agreement No. 679058 – DIRECT-fMRI).

The authors also thank Dina Arengoth and Pia Nyborg Nielsen for expert technical assistance. AW and TO acknowledge financial support from Lundbeck Foundation, Danish Multiple Sclerosis Society, Independent Research Fund Denmark

The authors would also like to offer special thanks to Shemesh Lab members in Champalimaud Center for Unknown that provided their extensive help during the acquisition stages and, in particular, to Teresa Serrades Duarte, Daniel Nunes, Rui Simões and Cristina Chavarrías.

# X. Figure captions

1. Example of acquired raw data and a corresponding data fit. Subfigures (A) and (B) show an example of a raw signal image acquired for low diffusion weighting ( $b = 0.2\,\mu m\,ms^{-2}$) in mid thoracic and low thoracic segments of a control sample. Subfigure (C) shows a high grade sample, with a visible lesion acquired with the same low diffusion weighting ( $b = 0.2\,\mu m\,ms^{-2}$). Data and data fit that correspond to three different voxels in the slice denoted in (C) are shown in subplot (D). Lesion voxel location is marked in red, NAWM voxel in green and GM voxel in magenta in subplot (C). Multiple data points plotted under each b-value on the x-axis correspond to different directions. The inset shows the part of the graph corresponding to $b=0.5\,\mu mms^{-2}$ enlarged.

2. Example outcome of lesion identification in 4 spinal cords in a lumbar segment. From left to right the grades are control, low grade, intermediate grade and high grade of EAE. For each of two subplots, the upper image represents a raw $T_2$-map, while the lower image shows the same map with the manual lesion delineation superimposed in yellow.

3. Correspondence between the grade of animal disability and lesion load in WM of spinal cord tissue. The bar plot shows the lesion load measured by relative volume (number of voxels in lesions divided by number of voxels in the WM of corresponding segment) averaged per sample. Different colors represent distinct grades of EAE. The bar plots from left to right correspond to mid-thoracic, lower-thoracic, lumbar segments of spinal cord. The last barplot represents average over all the segments. Each voxel corresponds to $6.1 \cdot 10^{-4} mm^3$, an average segment volume is ~2000-3000 voxels for mid thoracic, ~4000-5000 for lower thoracic and 6000-8000 for lumbar segments. Each bar plot is an average of 5 samples (low-grade), 3 samples (intermediate grade), 5 samples (high grade), 5 samples (control), 18 samples in total. Error bars depict standard deviation of values within samples. In a row under each one of the bars number of slices in particular segment and belong to a particular disability grade is provided. The legend provides number of animals in each of the groups.

4. Examples of parameter maps for each of the measured parameters in mid-thoracic segments of spinal cord. Each column (from left to right) corresponds to different grades of EAE disability: control animal, low grade, intermediate grade and high grade of EAE. Rows correspond to different measured parameters (A) (from top to bottom): mean diffusivity, MKT, FA, axial diffusivity, radial diffusivity, radial kurtosis, parallel kurtosis; (B): axonal water fraction, axonal diffusivity, axial extra-axonal diffusivity, radial extra-axonal diffusivity and concentration parameter of Watson distribution, the upper row depicts the delineation of spinal cord on the background of FA map

5. Examples of parameter distributions in the post-hoc parameter analysis for MKT in GM (A) and $K_\perp$ in NAWM (B) illustrated with box-plots. Each of 4 subplots corresponds to one of the spinal cord segments (mid-thoracic, low-thoracic and *lumbar and a graph for all voxels pooled across segments). Each box-plot represents parameter distribution for a corresponding EAE-grade (control animals, low-grade, medium grade, high grade). Blue dots correspond to the parameter means within each spinal cord. Asterisk denotes significant group-wise difference between the spinal cord means, tested with ANOVA as described in post-hoc analysis (C) illustrates the difference between the NAWM and lesion tissue in $K_\perp$. Each box-plot represents the distribution of values inside and outside the hyperintensity lesions. Blue dots correspond to the parameter means within each spinal cord. Asterisk denotes significant group-wise difference between the spinal cord means. In all three plots the central mark indicates the data median, the bottom and top edges indicate the 25th and 75th percentiles. The whiskers extend to the most extreme data points excluding outliers, and the outliers (voxels) are plotted individually in red.*

6. *Application of a "hybrid" biomarker (Eq. 6) on animal-wise data. From left to right the values are group-wise means of described control, low-grade, intermediate and high grade EAE. Error bars depict standard deviation of animal-wise estimates of the biomarker. Asterisk denotes statistical significance measured with 1-way ANOVA.*

# XI. Table captions

1. *Fit results of all measured parameters averaged for disability group. For each parameter a mean estimate provided along with standard deviation of error inside and outside the lesion (in NAWM). The mean and standard deviation of error was calculated only in the tissues in which a particular parameter was used for a successive LME analysis. Thus the statistics for MKT and MD was estimated only in GM, and the statistics for the rest of the DKI and WMM parameters was estimated only in WM. Note also that GM parameters (MD,MKT) were not calculated inside the lesions.*

2. *The results of the fit of the linear mixed effects model. For each of the studied parameters (in rows), the following are presented in columns: percent of outlier values removed, quality of LME fit $R^2_\beta$ (Edwards et al., 2008), p-values for coefficients of grade, lesion, segment and segment\*lesion, partial $R^2_\beta$(Edwards et al., 2008) of the same 4 coefficients and the results of the FDR multiple comparison test. Since lesions were registered only in WM, the coefficients of lesion are absent in GM .*

3. *Post-hoc analysis of parameters in GM and NAWM. Average value for all the NAWM or GM in the particular segment in each sample. (A) For each of the disability groups comparisons low-grade vs intermediate grade, low-grade vs high grade and intermediate grade vs high grade parameters that were found significant (p<0.05) after ANOVA of per-sample mean in each one of the segments is provided in the corresponding cell. An FDR correction of multiple comparisons has been taken into account. GM parameters are underlined. (B) Each of disability groups (low, intermediate, high) compared with the control group. Parameters that were found significant (p<0.05) after ANOVA of per-sample mean in each of the segments are listed in corresponding cells. An FDR correction has been performed. GM parameters are underlined.*

4. *Post-hoc analysis of NAWM against lesion tissue. For each WM parameter (in columns) we present the p-values that characterize the significance of difference in per-sample parameter mean inside against outside the lesion.*